\documentstyle[11pt, epsfig]{article}

\def\journalfont{\it}
\def\jou#1{{\journalfont #1\ }}

\def\CQG{\jou{  Class.\ Quantum Grav.}}

\def\beq{\begin{equation}}  \def\eeq{\end{equation}}
\def\beqa{\begin{eqnarray}} \def\eeqa{\end{eqnarray}}

\def\fr#1#2{{\textstyle{#1\over#2}}}    

\def\tr{\rm tr}

\def\R{{\bf R}}
 
\begin{document}

\title{%
   Dynamics of spatially homogeneous locally rotationally symmetric
   solutions of the Einstein-Vlasov equations}

      \author{A. D. Rendall \\
        {\it Max-Planck-Institut f\"ur Gravitationsphysik, 
        Am M\"uhlenberg 1}\\
        {\it D-14476 Golm, Germany}\\
        and \\ \\
       C. Uggla\\
       {\it Department of Physics, 
       University of Karlstad}\\
       {\it S--65188 Karlstad, Sweden} }
\maketitle

\begin{abstract}
The dynamics of the Einstein-Vlasov equations for a class of cosmological
models with four Killing vectors is discussed in the case of massive 
particles. It is shown that in all models analysed the solutions with
massive particles are asymptotic to solutions with massless particles 
at early times. It is also shown that in Bianchi types I and II the
solutions with massive particles are asymptotic to dust solutions at 
late times. That Bianchi type III models are also asymptotic to dust 
solutions at late times is consistent with our results but is not 
established by them. 

\end{abstract}



\section{Introduction}
The most popular matter content by far in the study of spatially 
homogeneous cosmological models is a perfect fluid with linear equation 
of state (see e.g., the book \cite{waiell97}). It is important to
know if the results obtained for this class are structurally stable if 
we change the matter content. Thus it is of interest to investigate other 
types of sources. Here we will consider certain diagonalizable
locally rotationally symmetric (LRS) spatially homogeneous models with 
collisionless matter. This class of models was previously studied in the 
case of massless particles in \cite{rentod99}. Here we will focus on the 
case with massive particles. 
We will recast Einstein's field equations into a form so that one part of
the boundary of the state space for the massive case can be identified with
the state space for the massless case while another part can be identified
with the state space for the corresponding dust equations. (In addition 
other parts of the boundary have the interpretation of state spaces 
associated with certain models with distributional matter.)
It will be shown that these boundary submanifolds are intimately connected
with the early and late time behaviour of the LRS massive collisionless gas
models respectively.

The results of our analysis can be summarized as follows. Consideration
is restricted to models of Bianchi types I, II and III. This is enough to 
display a large variety of phenomena. At early times, i.e. close to the
initial singularity, the dynamics of solutions with massive particles 
mimics closely the dynamics for the corresponding symmetry type with 
massless particles. In particular there are solutions whose behaviour
near the singularity is quite different from that of any fluid model 
of any of these Bianchi types. At late times, i.e. in a phase of 
unlimited expansion,
the general picture is that the dynamics resembles that of a dust model.
This is proved for Bianchi types I and II. For type III the results are
consistent with dust-like asymptotics but we were not able to prove that
this is what happens. If kinetic theory with massive particles always 
behaved like dust at late times this would provide a justification of the 
use of a fluid model in that regime.

The outline of the paper is as follows. In section 2 we derive 
the dynamical system. Sections 3, 4 and 5 analyse the models of types I, II
and III respectively, with the main results being stated in Theorems 2.1,
3.1 and 4.1. In section 6 we conclude with some remarks and speculations.
An appendix contains some information about dynamical systems which is
applied frequently in the paper.

\section{A dynamical systems formulation}

We will consider LRS models for which the metric can be written
in the form
\beq\label{eq:metric}
   ds^2 = -dt^2 + g_{11}(t)(\theta^1)^2 + 
           g_{22}(t)((\theta^2)^2 + (\theta^3)^2)\ ,
\eeq
where $\theta^i$ are suitable one-forms describing the various symmetry types.
The energy-momentum tensor $T_{ij}$ for the Einstein-Vlasov system with 
massive particles is assumed to be diagonal and is described by
\beqa\label{eq:rho}
  \rho &=& \int f_0(v_i)
  (m^2 + g^{11}(v_1)^2 + g^{22}((v_2)^2 + (v_3)^2))^{1/2}
  (\det g)^{-1/2} dv_1dv_2dv_3\ ,\nonumber \\
  p_i &=& \int f_0(v_i)g^{ii}(v_i)^2
  (m^2 + g^{11}(v_1)^2 + g^{22}((v_2)^2 + (v_3)^2))^{-1/2}
  (\det g)^{-1/2} dv_1dv_2dv_3\ ,
\eeqa 
where $\rho$ is the energy density and $p_i=T^i{}_i$ the pressure components 
of the energy-momentum tensor. The function $f_0$ is determined at some fixed
time $t_0$ by $f_0(v_i)=f(t_0,v_i)$ where $f$ is the phase space density
of particles. The covariant components $v_i$ are independent of time.
The function $f_0$ satisfies the condition
$f_0(v_1,v_2,v_3) = F(v_1,(v_2)^2 + (v_3)^2)$. 

Some further technical conditions will be imposed on $f_0$. It is assumed to 
be non-negative and have compact support. It is also assumed that the 
support does not intersect the coordinate planes $v_i=0$. A function $f_0$
with this property will be said to have split support. The reason for the
assumption of split support will be seen later. In the following it will
always be assumed without further comment that the data considered have
split support. It follows from the assumptions already made that 
$f_0(x_i)=f_0(-x_i)$ for $i=2,3$. It will
be assumed that this also holds for $i=1$ and functions $f_0$ with this
property will be called reflection-symmetric. This ensures that the form
of the phase space density of particles is compatible with a diagonal metric
and, in particular, that the energy-momentum tensor is diagonal. For 
the symmetry types to be considered in the following it then follows that
the entire system consisting of geometry and matter is invariant under
three commuting reflections. For this reason, solutions where the metric
is diagonal and $f_0$ has the symmetry properties just mentioned will be
called reflection-symmetric. A solution is said to be isotropic if 
$f_0(v_1,v_2,v_3) = F((v_1)^2 + (v_2)^2 + (v_3)^2)$ and if
$g_{11} \propto g_{22} \propto g_{33}$ for all time.

The momentum constraints are automatically satisfied for these models.
Only the Hamiltonian constraint and the evolution equations are left.
Instead of considering a set of second order equations in terms of e.g.,  
$a$ and $b$, where 
\beq
  a^2 = g_{11}\ ,\quad b^2 = g_{22}\ ,
\eeq
we will reformulate these equations as a first order system of ODEs by 
introducing a new set of variables. The mean curvature ${\tr} k$
(where $k_{ij}$ is the second fundamental form) is given by
\beq
  {\tr} k = -(a^{-1}da/dt + 2b^{-1}db/dt)\ .
\eeq
A new dimensionless
time coordinate $\tau$ is defined by $-\fr13\int_{t_0}^t {\tr} k(t)dt$
for some arbitrary fixed time $t_0$. (We will follow the conventions
in \cite{waiell97}. The time variable thus differs by a factor 3 from the 
one in \cite{rentod99}). In the following a dot over a quantity denotes 
its derivative with respect to $\tau$. 
The Hubble variable $H$ is given by $H = -{\tr} k/3$.
Now define the following dimensionless variables:
\beqa\label{eq:var}
  z &=& m^2/(a^{-2} + 2b^{-2} + m^2)\ ,\nonumber \\
  s &=& b^2/(b^2 + 2a^2)\ ,\nonumber \\
  M_2 &=& \sigma_2 (a^2/b^4)({\tr} k)^{-2}\ ,\nonumber \\
  M_3 &=& 3\sigma_3 b^{-2}({\tr} k)^{-2}\ ,\nonumber \\
  \Sigma_+ &=& -3(b^{-1}db/dt)({\tr} k)^{-1} - 1\ ,
\eeqa 
where $\sigma_2$ is 1 for Bianchi types II, VIII, IX and
0 for Bianchi types I, III and the Kantowski-Sachs (KS) models.
The coefficient $\sigma_3$ is $1$ for types III and VIII. It is
$-1$ for KS and type IX and 0 for types I, II
\footnote{The motivation for the variable $s$ comes from more general diagonal
models where it is convenient to introduce variables of the type
$s_i= g^{ii}/(g^{11} + g^{22} + g^{33})$. $s$ is simply $s_1$ in the case 
when $g^{22}=g^{33}$.}. 
These variables lead to a 
decoupling of the equation for the only remaining dimensional variable
$H$ (or equivalently ${\tr} k$)
\beq
  \dot{H} = -(1 + q)H\ ,
\eeq
where the deceleration parameter $q$ is given by
\beq\label{eq:q}
  q = 2\Sigma_+^2 + \fr12 \Omega (1 + R)\ .
\eeq
The quantity $R$ is defined by
\beq\label{eq:R}
  R = (p_1 + 2p_2)/\rho\ ,
\eeq 
where
\beqa\label{eq:p}
  p_1/\rho &=& (1 - z)sg_1/h\ ,\nonumber \\
  p_2/\rho &=& \fr12 (1 - z)(1 - s)g_2/h\ ,\nonumber \\
  g_{1,2} &=& \int f_0(v_i) (v_{1,2})^2 
  [z + (1 - z)(s (v_1)^2 + \fr12 (1 - s)((v_2)^2 + (v_3)^2))]^{-1/2}
  dv_1dv_2dv_3\ ,
  \nonumber \\
  h &=& \int f_0(v_i) 
          [z + (1 - z)(s (v_1)^2 + \fr12 (1 - s)((v_2)^2 + (v_3)^2))]^{1/2}
          dv_1dv_2dv_3\ .
\eeqa

The assumption of split support ensures that the function $R(s,z)$ is a
smooth ($C^\infty$) function of its arguments. The related quantity $R_+$ 
defined by
\beq\label{eq:rp}
R_+ = (p_2 - p_1)/\rho\ .
\eeq
is a smooth function of $s$ and $z$ for the same reason.

The normalized energy density $\Omega=\rho/(3H^2)$ is determined by 
the Hamiltonian constraint and, in units where $G=1/8\pi$, is given by
\beq\label{eq:om}
  \Omega = 1 - \Sigma_+^2 - M_2 - M_3\ .
\eeq 

The assumption of a distribution of massive particles with non-negative 
mass leads to inequalities for $R$, $R_+$ and $\Omega$. Firstly,
$0\le R\le 1$ with $R=0$ only when $z=1$ and $R=1$ only when $z=0$.
Secondly, $-R\le R_+\le \fr12 R$ with $R_+=\fr12 R$ for $s=0$ and $R_+=-R$ 
for $s=1$. Thirdly $\Omega\ge 0$. Using these inequalities in equation 
(\ref{eq:q}) in turn results in $0\leq q \leq 2$ for Bianchi types I, II,
III and VIII (i.e., the same inequality as for causal perfect fluids, see 
\cite{waiell97}).
 
The remaining dimensionless coupled system is:
\beqa\label{eq:eq1}
  \dot{\Sigma}_+ &=& -(2-q)\Sigma_+ - {\cal S}_+ + \Omega R_+\ ,\nonumber \\
  \dot{s} &=& 6s(1 - s)\Sigma_+\ ,\nonumber \\
  \dot{z} &=& 2z(1 - z)(1 + \Sigma_+ - 3\Sigma_+s)\ ,\nonumber \\
  \dot{M}_2 &=& 2(q - 4\Sigma_+)M_2\ ,\nonumber \\
  \dot{M}_3 &=& 2(q - \Sigma_+)M_3\ ,
\eeqa 
where ${\cal S}_+$ is given by
\beq
{\cal S}_+ = -4M_2 - M_3.
\eeq 

\begin{figure}[ht]
  \begin{center}
    \epsfig{figure=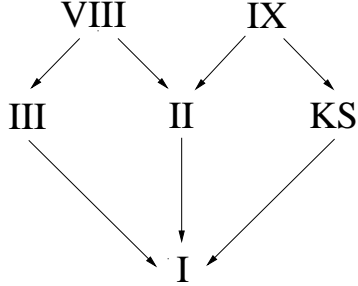, width=0.3\textwidth}
    \caption{Symmetry reduction diagram for the diagonal LRS models.}
    \label{fig:symred}
  \end{center}
\end{figure}

There are a variety of submanifolds corresponding to different symmetry types:
\beqa\label{eq:sym}
   S_{\rm I} &:& M_2 = M_3 = 0\ ,\nonumber \\
   S_{\rm II} &:& M_2 > 0\ , M_3 = 0\ ,\nonumber \\
   S_{\rm III} &:& M_2 = 0\ , M_3 > 0\ ,\nonumber \\
   S_{\rm KS} &:& M_2 = 0\ , M_3 < 0\ ,\nonumber \\
   S_{\rm VIII} &:& M_2 > 0\ , M_3 > 0\ , (1 - s)M_3 = 6M_2 s\nonumber \\ 
   S_{\rm IX} &:& M_2 > 0\ , M_3 < 0\ , (1 - s)M_3 = -6M_2 s\ .
\eeqa 
The relationship between the various models can be visualized in a symmetry 
reduction diagram given in Fig.\ \ref{fig:symred} (a collective treatment 
of the corresponding vacuum models from a Hamiltonian perspective and with the
aim of quantizing the models was given in \cite{ash93}). Note that while this
diagram accurately reflects the relationship of the geometry in the
different cases, the relationship of the matter content is more subtle 
when types VIII or IX are involved. This complication does not occur for 
the Bianchi types studied in detail in this paper and will therefore not be 
discussed further here.

Note that a non-negative energy density implies that $\Omega \geq 0$, which in
turn implies that our variables are bounded for types I,II,III and VIII,
since $M_2$ and $M_3$ are non-negative and
since by definition $z$ and $s$ are bounded. These models expand
indefinitely. The KS and type IX models are recollapsing models and since 
$H$ becomes zero at the point of maximal expansion, the Hubble-normalized 
variables blow up at this point. However, one can find other variables that
are bounded along the lines found in \cite{uggzur90}.
Neither are the above variables `optimal' for the other LRS models. One can 
adapt to the particular mathematical features these models exhibit. 
However, we choose to use the above formulation since 
the present variables are easier to interpret physically and
are naturally generalizable to more general non-LRS models.
For simplicity we will from now on study Bianchi types I,II, and III.

It is of interest to note that the metric functions $a,b$ are expressible 
in terms of $s,z$ in the massive case. The relations are
\beq
  a^2 =  z(m^2s(1-z))^{-1},\quad
  b^2 = 2z(m^2(1-s)(1-z))^{-1}\ .
\eeq

In addition to the symmetry submanifolds there are a number of other
boundary submanifolds:
\beqa\label{eq:sub}
   z &=& 0,1\ ,\nonumber \\
   s &=& 0,1\ ,\nonumber \\    
   \Omega &=& 0\ .
\eeqa 
The submanifold $z=0$ corresponds to the massless case.
The submanifold $z=1$ leads to a decoupling of the
$s$-equation, leaving a system identical to the corresponding 
dust equations. 
The submanifolds $s = 0,s = 1$ correspond to problems with 
$f_0$ being a distribution while $\Omega = 0$ constitutes the vacuum 
submanifold with test matter.
Apart from these solutions there exists an isotropic solution in Bianchi 
type I characterized by $\Sigma_+ = R_+ = 0$ and a constant value for $s$ 
that depends on the function $f_0$.

Including these boundaries yields compact state spaces for types I,II and III.
In order to apply the standard theory of dynamical systems the coefficients
must be $C^1$ on the entire compact state space $G$ of a given model. 
This is necessary even for uniqueness. In the present case it
suffices to show that $R$ and $R_+$ are $C^1$ on $G$, i.e, 
that they are $C^1$ for $s,z$ when $0\leq s \leq 1\ ,0\leq z \leq 1$. 
As has already been pointed out, this follows from the assumption of
split support, which even implies the analogous statement with $C^1$
replaced by $C^\infty$. It would be possible to get $C^1$ regularity
under the weaker assumption that $f_0$ vanishes as fast as a sufficiently
high power of the distance to the coordinate planes. We have not, however,
examined in detail how large the power would have to be since this is of
little relevance to our main concerns in this paper.

Of key importance is the existence of a monotone function in the 
`massive' interior part of the state space:
\beqa\label{eq:mon}
   M &=& (s(1-s)^2)^{-1/3}z(1 - z)^{-1}\ ,\nonumber \\
   \dot{M} &=& 2M\ .
\eeqa 
Note that the volume $ab^2$ is proportional to $M^{3/2}$. 
This monotone function rules out any interior $\omega$- and $\alpha$-limit
sets and forces these sets to lie on the $s=0$, $s=1$, $z=0$ or $z=1$ parts of 
the boundary.

\section{Type I models}

It is natural to start investigating the type I system since it is a
submanifold of the state space of all other symmetry types. 
The physical state space, $G$, of these models is given by the region in
$\R^3$ defined by the inequalities $-1\le \Sigma_+
\le 1$, $0\le s\le 1$ and $0\le z\le 1$.

To understand the
dynamics of the type I models, it is necessary to determine the stationary 
points and their stability.
The coordinates, in terms of $(\Sigma_+,s,z)$, of the various stationary
points are the following: $(0,s_0,0)$,$(\fr12 ,0,0)$,
$(1,0,0)$,$(1,1,0)$,$(-1,1,0)$,$(1,0,1)$,$(1,1,1)$,$(-1,1,1)$,
where $s_0$ is a particular constant value of $s$ depending on the function 
$f_0$ (see \cite{rentod99}). These points are called $P_1,...P_8$.
(Note that they are numbered differently in the massless case compared to 
those in \cite{rentod99}.)
In addition there exist two lines of equilibrium points, 
$(-1,0,K),(0,F,1)$, denoted by $L_1,L_2$, where $K$ and $F$ are constant
values. 
The points $P_1,P_2,P_4,P_6,P_7,P_8$ are hyperbolic saddles while 
$P_5$ is degenerate, with one zero eigenvalue. The point $P_3$ is a
hyperbolic source. The line $L_1$ is a transversally hyperbolic saddle
while the line $L_2$ is a transversally hyperbolic sink. (For an
explanation of this terminology we refer to the appendix.)

\begin{figure}[ht]
  \begin{center}
    \epsfig{figure=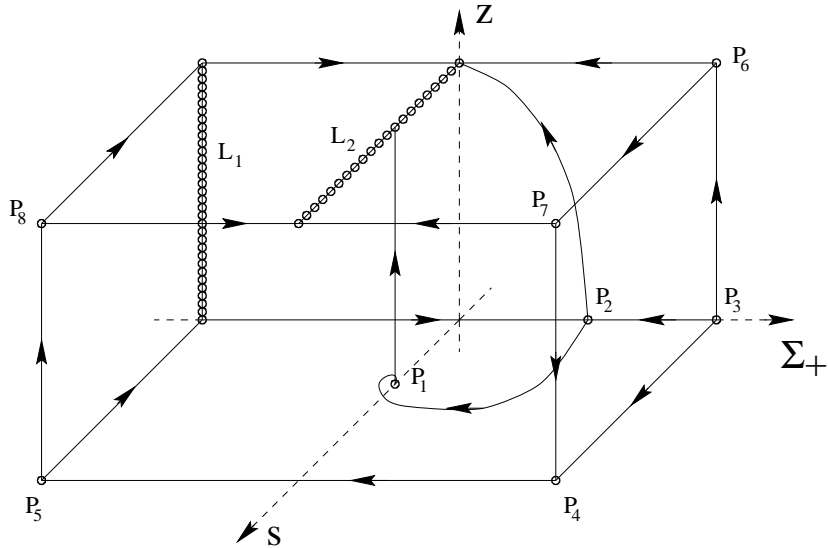, width=0.7\textwidth}
    \caption{The LRS type I state space together with equilibrium points 
             and separatrix orbits.}
    \label{fig:lrs1}
  \end{center}
\end{figure}

The state space together with equilibrium points and separatrix orbits
is depicted in Fig.\ \ref{fig:lrs1}.

The main result in this section is the following theorem:

\noindent
{\bf Theorem 3.1} If a smooth non-vacuum reflection-symmetric LRS
solution of Bianchi type I of the Einstein-Vlasov equations for massive 
particles is represented as a solution of (\ref{eq:eq1}) with $M_2=M_3=0$
then for $\tau \rightarrow \infty$ it converges to a point of the line $L_2$.
For $\tau \rightarrow -\infty$ there exists 

\noindent
(i) a single (isotropic) solution that converges to $P_1$ and

\noindent
(ii) a one-parameter set of solutions lying on the unstable manifold of 
$P_2$ and 

\noindent
(iii) all remaining solutions belong to
a two-parameter set (the generic case) of solutions converging to $P_3$.

This will be proved in a series of lemmas. We refer to 
\cite{waiell97,rentod99} for terminology from the theory of dynamical systems.

\noindent
{\bf Lemma 3.1} There exist open neighbourhoods $U_1$ and $U_2$ of 
the point $P_3$ and the line $L_2$  respectively such that:

\noindent
(i) if a solution belongs to $U_1$ at any time it belongs to $U_1$ at
all earlier times and its $\alpha$-limit set consists of the point $P_3$
alone. 

\noindent
(ii) if a solution belongs to $U_2$ at any time it belongs to $U_2$ at
all later times and its $\omega$-limit point consists of a single point
of the line $L_2$.

\noindent
{\bf Proof} Part (i) follows from the fact that $P_3$ is a hyperbolic 
source and the Hartman-Grobman theorem. Part (ii) follows from the
fact that $L_2$ is a transversally hyperbolic sink and the reduction
theorem (\cite{rentod99}, Theorem A1).

As a step towards analysing the dynamics of the full system we determine
the $\omega$-limit points of solutions of the dynamical system on the 
parts of the boundary of $G$ defined by $s=0$ and $s=1$. This information
will later be combined with the monotone function $M$ when determining the 
$\omega$-limit sets of solutions of the full system. In the case of the
$\alpha$-limit sets the monotone function alone accomplishes the same
thing.

\noindent
{\bf Lemma 3.2} A solution of the restriction of the system to the part of
the boundary of $G$ defined by $s=1$ for which neither $z$ nor $\Sigma_+$ 
take on one of their limiting values has the endpoint of $L_2$ with $s=1$ 
as its $\omega$-limit set.  

\noindent
{\bf Proof} If $\Sigma_+\ge 0$ at any time, then $\Sigma_+$ is decreasing.
The rate of decrease remains uniform as long as $\Sigma_+$ does not tend
to zero. It follows that after a finite time $\Sigma_+$ must be strictly 
less than $1/2$. On the other hand, $z$ is monotone increasing in the 
region $\Sigma_+<1/2$ and the rate of increase remains uniform as long as 
$z$ does not tend to one. It follows that $z\to 1$ as $\tau\to\infty$.
If $\Sigma_+$ tends to zero in this limit then the conclusion of the lemma
holds. Otherwise $\Sigma_+$ must become negative at some time. Thus it can 
be seen that any $\omega$-limit points satisfy $z=1$ and 
$-1\le\Sigma_+\le 0$. From part (ii) of Lemma 3.1 it follows
that any solution which enters $U_2$ has the desired $\omega$-limit set. 
Since the $\omega$-limit set is a union of orbits, it is possible as a
consequence to exclude the points with $z=1$ and $-1<\Sigma_+<0$ from the 
$\omega$-limit set. To complete the proof of the lemma it remains only to
exclude the point $P_8$ from the $\omega$-limit set. This point is a 
hyperbolic saddle of the restriction of the system to $s=1$ and so it
follows from the discussion in the appendix and what has been proved 
already that it cannot belong to the $\omega$-limit set. For if $P_8$
belonged to the $\omega$-limit set points of its stable and unstable
manifolds would also have to do so, and this has already been ruled
out.  

\noindent
{\bf Lemma 3.3}  A solution of the restriction of the system to the part of
the boundary of $G$ defined by $s=0$ for which neither $z$ nor $\Sigma_+$ 
take on one of their limiting values has the endpoint of $L_2$ with $s=0$ 
as its $\omega$-limit set. 

\noindent
{\bf Proof} Along any solution of this system $z$ is monotone increasing
on the part of the state space of the restricted dynamical system with
$\Sigma_+\ne -1$ and $z(1-z)\ne 0$. Hence, by the monotonicity principle 
(see \cite{waiell97}), any $\omega$-limit point must satisfy $z=1$ or
$\Sigma_+=-1$. However $\Sigma_+$ is increasing for $\Sigma_+$ close
to but not equal to $-1$. Hence there can be no $\omega$-limit points with 
$\Sigma_+=-1$. It follows that $z$ tends to one as $\tau\to\infty$ for 
any solution and any $\omega$-limit point satisfies $z=1$. From Lemma 3.1, 
any solution which enters $U_2$ has the desired $\omega$-limit set. Arguing 
as in the proof of Lemma 3.3 allows points with $-1<\Sigma_+<0$ and 
$0<\Sigma_+<1$ to be excluded. The point $P_6$, which is a hyperbolic 
saddle of the restricted system, can be eliminated in the same way as was 
done in the case of $P_8$ in the proof of Lemma 3.2 using the results of 
the discussion in the appendix. Finally, the non-existence of $\omega$-limit
points with $\Sigma_+=-1$, already mentioned above, shows that the endpoint 
of the line $L_1$ cannot lie in the $\omega$-limit set.

\noindent
{\bf Lemma 3.4} If a solution lies in the interior of $G$, then unless it 
lies on the unstable manifold of $P_1$ or $P_2$ its $\alpha$-limit set 
consists of the point $P_3$.

\noindent
{\bf Proof} Consider a solution in the interior of $G$ which does not lie
on the unstable manifold of $P_1$ or $P_2$. If it intersects $U_1$ then by
Lemma 3.1 its $\alpha$-limit set consists of the point $P_3$. There can be no 
other $\alpha$-limit points in $U_1$. Because the function $M$ tends to zero
along the solution as $\tau\to -\infty$ the $\alpha$-limit set must be 
contained in the surface $z=0$. Recall that the surface $z=0$ corresponds to
the case of massless particles which was analysed completely in 
\cite{rentod99}.
(Note that the stationary points were numbered differently in that paper.)
Consider the boundary of the surface $z=0$. Arguing as in the proof of
Lemma 3.2, the lines joining $P_3$ to $P_2$ and $P_4$ can be excluded from
the $\alpha$-limit set. The discussion of the appendix and the fact that 
$P_4$ is a hyperbolic saddle with stable manifold $\Sigma_+=1$ and unstable 
manifold the line connecting $P_4$ to $P_5$ can be used to exclude that 
line and the point $P_4$ itself. The line connecting $P_5$ to the endpoint 
of the line $L_1$ can be excluded in an analogous way, noting that the 
non-hyperbolic point $P_5$ is also covered by the discussion of the appendix.
The point $P_5$ is also excluded by this argument. Applying the reduction 
theorem allows the line joining the endpoint of the line $L_1$ to $P_2$ to 
be excluded together with the endpoint of $L_1$. At this stage we can also
exclude the point $P_2$ itself, using the results of the appendix  again and 
the fact that
by assumption the solution does not lie on the unstable manifold of $P_2$. 
Thus the only point of the boundary of the set $z=0$ which can belong to the
$\alpha$-limit set is $P_3$. Now suppose that a point of the interior of the 
surface belongs to the $\alpha$-limit set. If it is a point of the unstable
manifold of $P_2$ then $P_2$ also belongs to the $\alpha$-limit set, in
contradiction to what has just been proved. If it is some other point other
than $P_1$ then, using the fact that the $\alpha$-limit set is a 
union of orbits and Theorem 3.1 of \cite{rentod99}, it follows that $P_3$ 
belongs to
the $\alpha$-limit set and we obtain a contradiction again. Finally, if it
were $P_1$ then the results of the appendix would imply that other points of 
the interior would belong to the $\alpha$-limit set, and this has just been 
ruled out.  

\noindent
{\bf Lemma 3.5} The $\omega$-limit point of each solution in the interior
of $G$ is a point of the line $L_2$.

\noindent
{\bf Proof} Note first that the function $M$ goes to infinity along any 
such solution as $\tau\to\infty$. It follows that any $\omega$-limit
point must satisfy $z=1$, $s=0$ or $s=1$. If the solution enters the set
$U_2$ then by part (ii) of Lemma 3.1 the $\omega$-limit set is as claimed.
There are no other $\omega$-limit points of any solution in $U_2$.
Consider now the evolution of $\Sigma_+$ on the surface $z=1$. It either
increases from $-1$ to $0$ or decreases from $1$ to $0$. Since
the $\omega$-limit set is a union of orbits, we conclude
that no point of the interior of the surface $z=1$ or its boundary lines
$s=0$ and $s=1$ other than the points of the line $L_2$ can belong to the
$\omega$-limit set. Using once more the fact that the $\omega$-limit set is
a union of orbits, it is possible to exclude the interior of the surface
$s=1$ from the $\omega$-limit set by Lemma 3.2 and the interior of $s=0$ by 
Lemma 3.3. Now all remaining possibilities other than points on $L_2$ will be
excluded successively. The nature of the line $L_1$ as a transversely
hyperbolic saddle suffices to eliminate it, as well as the lines joining 
it to $P_8$ and $P_2$. The point $P_3$, being a hyperbolic source, is 
clearly ruled out, and with it the lines joining it to $P_2$ and $P_6$.
Further applications of the results of the appendix rule out 
the remaining lines, namely those joining $P_8$ to $P_5$, $P_5$ to $P_4$, 
$P_4$ to $P_7$ and $P_7$ to $P_6$. It follows that the $\omega$-limit set
is contained in the line $L_2$. Applying the reduction theorem then shows
that the $\omega$-limit set is a single point of $L_2$.

\vskip 10pt\noindent
The results of Lemma 3.4 and Lemma 3.5 together imply Theorem 3.1.

Theorem 3.1 has been formulated entirely in terms of the dynamical systems
picture. It should, however, be pointed out that this allows asymptotic
expansions for all quantities of geometrical or physical interest near 
the singularity or in an expanding phase to be obtained if desired. For 
example, in an expanding phase in type I the following expansions can be 
derived:
\beqa
\Sigma_+&=&\alpha t^{-1}+o(t^{-1})         \\
s&=&s_0-\fr49 s_0(1-s_0)t^{-1}+o(t^{-1})   \\
z&=&1-\beta t^{-4/3}+o(t^{-4/3})           \\
H&=&\fr23 t^{-1}+O(t^{-7/3})               \\ 
\Omega&=&1-\alpha t^{-2}+o(t^{-2})           \\
\rho&=&\fr49 t^{-2}-\fr43\alpha^2 t^{-4}+o(t^{-4})  \\
p_1&=&O(t^{-10/3})
\eeqa
Here $\alpha$ and $\beta$ are constants depending on the solution.
It should be emphasized that these are not just formal expansions, but
rigorous results which emerge from the dynamical systems analysis.

A particular consequence of Theorem 3.1 is that all LRS type I models 
isotropize at late times. This was already proved by other means in 
\cite{rendall96}, where it was also shown that non-LRS models of Bianchi 
type I isotropize and have dust-like behaviour for $\tau\to\infty$.

\section{Type II models}

The physical state space, $G$, of the LRS type II models is given by the 
region in $\R^4$ defined by the inequalities $M_2\ge 0$, 
$0\le s\le 1$, $0\le z\le 1$, and $1 - \Sigma_+^2 - M_2 \geq 0$.

The coordinates, in terms of $(\Sigma_+,s,z,M_2)$, of the various stationary
points are the following: $(0,s_0,0,0)$,$(\fr12 ,0,0,0)$,
$(1,0,0,0)$,$(1,1,0,0)$,$(-1,1,0,0)$,$(1,0,1,0)$,$(1,1,1,0)$,$(-1,1,1,0)$,
\newline
$(\fr15 ,1,0,\fr{6}{25})$,$(\fr18 ,0,1,\fr{3}{64})$,$(\fr18 ,1,1,\fr{3}{64})$,
where $s_0$ is the same particular constant value of $s$ that appeared in the
previous type I section. These points are called $P_1,...P_{11}$
(note that they are numbered differently than in \cite{rentod99}, 
in the massless case).
In addition there exist two lines of equilibrium points, 
$(-1,0,K,0),(0,F,1,0)$, denoted by $L_1,L_2$, where $K$ and $F$ are 
constants. The first eight stationary points and the two lines
correspond to points and lines of the same name in the Bianchi I system
and their coordinates are obtained by appending a zero to those of the 
Bianchi I points.  
The points $P_1,P_2,P_3,P_4,P_6,P_7,P_8,P_9,P_{10}$ 
are hyperbolic saddles while 
$P_5$ is degenerate, with one zero eigenvalue. The point $P_{11}$ is a
hyperbolic sink with two real and two complex eigenvalues. 
The lines $L_1$ and $L_2$ are transversally hyperbolic saddles.

To prove results about the global properties of solutions it is helpful
to use certain monotone functions. The first is defined for $s<1$ by
\beq
Z_1=(2s/(1-s))^{4/3}M_2
\eeq
This is obtained by rewriting the function whose time derivative was 
calculated in equation (23) of \cite{rentod99} in terms of the variables
of this paper and observing that it remains monotone in the massive case.
It satisfies $\dot Z_1=2q Z_1$. The second is obtained by combining $Z_1$
with the monotone function $M$ available for all the Bianchi types 
considered in this paper. Let $Z_2=Z_1 M^{-2}=2^{4/3}s^2M_2z^{-2}(1-z)^2$ 
for $z>0$. It satisfies $\dot Z_2=2(q-2)Z_2$.  The function $Z_1$ is 
defined on the part of the Bianchi II state space where $s\ne 1$ and 
monotonically increasing except where it vanishes. This is clear if $q\ne 0$. 
If $q=0$ it follows that $\Sigma_+=0$ and $M_2=1$ and at points satisfying 
these conditions $\dot\Sigma_+\ne 0$. The function $Z_2$ is defined
on the part of the Bianchi II state space where $z>0$ and is monotonically 
decreasing except on the set where it vanishes, since $q=2$ implies
$Z_2=0$.

\noindent
{\bf Theorem 4.1} If a smooth non-vacuum reflection-symmetric LRS
solution of Bianchi type II of the Einstein-Vlasov equations for massive 
particles is represented as a solution of (\ref{eq:eq1}) with $M_3=0$,
then for $\tau \rightarrow \infty$ it converges to $P_{11}$.
For $\tau \rightarrow -\infty$ there exists 

\noindent
(i) a one-parameter set of solutions converging to the unstable 
manifold of $P_1$ and 

\noindent
(ii) a three-parameter set of all remaining solutions converging to the 
heteroclinic cycle on the $z=0$ submanifold, consisting of the orbits 
connecting the $z=0$ endpoint of the line $L_1$ to $P_5$, $P_5$ to $P_4$,
$P_4$ to $P_3$ on the type I boundary and $P_3$ to 
the $z=0$ endpoint of the line $L_1$ via the vacuum boundary.

\noindent
{\bf Lemma 4.1} If a solution belongs to the interior of the type II state
space then any $\alpha$-limit point satisfies $z=0$ and $sM_2=0$. Any 
$\omega$-limit point satisfies $s=1$ and $(z-1)M_2=0$. 

\noindent
{\bf Proof} From the evolution equation for $M$ it follows that $z=0$
for any $\alpha$-limit point and that for any $\omega$-limit point 
$z=1$, $s=0$ or $s=1$. Next the monotonicity principle will be applied to 
the functions $Z_1$ and $Z_2$. Applying it to $Z_1$ on the region where
$Z_1\ne 0$ shows that for any $\alpha$-limit point $s=0$ or $M_2=0$. It also
shows that there are no $\omega$-limit points with $s=0$. Combining this 
with the information obtained already shows that any $\omega$-limit point 
satisfies $z=1$ or $s=1$. If $z\ne 1$ then it follows from the 
monotonicity principle applied to $Z_2$ that $M_2=0$ for any 
$\omega$-limit point. The monotonicity of $Z_1$ then implies that 
$s\to 1$ as $\tau\to\infty$.

\noindent
{\bf Lemma 4.2} Consider the dynamical system obtained by restricting the
type II system to the plane defined by the conditions $s=1$ and $z=1$. If 
a solution belongs to the interior of the state space for this restricted 
system then it converges to $P_{11}$ as $\tau\to\infty$.

\noindent
{\bf Proof} The restricted dynamical system is identical with that
for type II dust solutions. In \cite{waiell97} it was proved
by using a monotone function derived by Hamiltonian methods that for
$\tau\to\infty$ the dust solutions satisfy $\Sigma_+\to \fr18$ and 
$M_2\to \fr3{64}$. Hence it can be concluded that the solution
approaches $P_{11}$ as $\tau\to\infty$.

\noindent
{\bf Lemma 4.3} If a solution lies in the interior of the type II state
space then unless it lies on the unstable manifold of $P_1$ (and this
does occur) the $\alpha$-limit set consists of the heteroclinic cycle 
described in the statement of Theorem 4.1.

\noindent
{\bf Proof} By Lemma 4.1 we know that any $\alpha$-limit point satisfies
$z=0$. Moreover it satisfies $M_2=0$ or $s=0$. The situation is very 
similar to that in the massless case treated in \cite{rentod99} and 
the proof may be taken over rather directly. It is only necessary to
pay attention to the fact that it is the nature of the stationary points 
in the full massive Bianchi II state space which must be taken into account 
and that the notation is different.

Suppose that the $\alpha$-limit set contains a point with $s=0$ and
$M_2\ne 0$. Then by Lemma 4.3 of \cite{rentod99} it contains the 
endpoint of $L_1$ and either $P_2$ or $P_3$. On the other hand, if 
it contains a point with $M_2=0$ then this belongs to the massless
Bianchi I state space. Then it must contain one of the points $P_1$, 
$P_2$, $P_3$, $P_4$, $P_5$ or the endpoint of $L_1$. To prove the lemma
we may assume that the solution does not lie on the unstable manifold 
of $P_1$. If $P_1$ nevertheless belonged to the $\alpha$-limit set
then this set would have to include points belonging to the unstable 
manifold of $P_1$ other than $P_1$ itself. But these satisfy neither 
$s=0$ nor $M_2=0$ and so this gives a contradiction. Thus under the
given assumptions the $\alpha$-limit set does not contain $P_1$.
If the $\alpha$-limit set contained a point with $M_2=0$, $|\Sigma_+|<1$
and $0<s<1$ it would contain $P_1$ (in its role as $\omega$-limit set
for Bianchi type I solutions), leading once more to a contradiction.
If the $\alpha$-limit set contained $P_2$ then by the results of the
appendix it would contain $P_1$, which is also not possible.
Applying Lemma 4.3 of \cite{rentod99} again allows points with $M_2\ne 0$ 
which are not on the vacuum boundary to be excluded from the $\alpha$-limit 
set. The straight lines joining $P_2$ to $P_3$ and the endpoint of $L_1$
are excluded as well. The conclusion is that the $\alpha$-limit set is
contained in the heteroclinic cycle mentioned in the statement of
Theorem 4.1. It remains to show that it is the whole heteroclinic cycle.
This is straightforward to do using the results of the appendix.

\noindent
{\bf Lemma 4.4} If a solution lies in the interior of the type II state
space then it converges to the point $P_{11}$ as $\tau\to\infty$.

\noindent
{\bf Proof} Consider any $\omega$-limit point with $z\ne 1$. Then by Lemma
4.1 this point satisfies $s=1$ and $M_2=0$. Any nearby $\omega$-limit points
must also satisfy these conditions. If any of these limit points satisfied
$z=0$ then $P_4$ and $P_5$ would be $\omega$-limit points of the given 
solution. Using the saddle point properties of these points then shows that
$P_7$ and $P_8$ belong to the $\omega$-limit set. Repeating the same 
argument shows that the endpoint of $L_1$ with $s=1$ is an $\omega$-limit
point. The fact that this point is a transversely hyperbolic saddle implies 
that its unstable manifold in the hyperplane $s=1$ is contained in the 
$\omega$-limit set. By Lemma 4.2 the $\omega$-limit set also contains
$P_{11}$. Since $P_{11}$ is a hyperbolic sink this contradicts the assumption
$z\ne 1$. Thus we conclude that the entire $\omega$ limit set is contained
in the plane defined by the equations $s=1$ and $z=1$. The argument just 
given rules out the possibility of $\omega$-limit points with $M_2=0$.
Applying Lemma 4.2 once more shows that the only possible $\omega$-limit 
point which does not lie on the vacuum boundary is $P_{11}$. Finally the
fact that $P_7$ and $P_8$ are hyperbolic saddles can be used to rule out
points of the vacuum boundary, thus completing the proof.

\section{Type III models}

The physical state space, $G$, of the LRS type III models is given by the 
region in $\R^4$ defined by the inequalities $M_3\ge 0$, 
$0\le s\le 1$, $0\le z\le 1$, and
$1 - \Sigma_+^2 - M_3 \geq 0$.

The coordinates, in terms of $(\Sigma_+,s,z,M_3)$, of the various stationary
points are the following: $(0,s_0,0,0)$,$(\fr12 ,0,0,0)$,
$(1,0,0,0)$,$(1,1,0,0)$,$(-1,1,0,0)$,$(1,0,1,0)$,$(1,1,1,0)$,$(-1,1,1,0)$,
\newline
$(\fr12 ,0,0,\fr34)$,$(\fr12 ,0,1,\fr34)$,
where $s_0$ the same particular constant value of $s$ that appeared in the
previous type I section. These points are called $P_1,...P_{10}$
(note that they are numbered differently than in \cite{rentod99}, in
the massless case).
In addition there exist three lines of equilibrium points, 
$(-1,0,K,0,0),(0,F,1,0,0),(\fr12,1,z_0,\fr34)$, denoted by $L_1,L_2,L_3$, 
where $K,F$ and $z_0$ are constants.
The first eight stationary points and the first two lines
correspond to points and lines of the same name in the Bianchi I system
and their coordinates are obtained by appending a zero to those of the 
Bianchi I points. 
The points $P_1,P_2,P_4,P_6,P_7,P_8,P_9$ are hyperbolic saddles while 
$P_5$ and $P_{10}$ are degenerate, with one zero eigenvalue each. 
The point $P_3$ is a hyperbolic source. 
The lines $L_1$ and $L_2$ are transversally hyperbolic saddles
while the line $L_3$ is degenerate with two zero and 
two negative eigenvalues.

To prove global results about the global properties of solutions
it is useful to note the 
existence of the following bounded monotone function
\beqa\label{eq:mon3}
   {\tilde M}_3 &=& M_3 (2 - \Sigma_+)^{-2}\ ,\nonumber \\
   \dot{{\tilde M}_3} &=& 
   2{\tilde M}_3[(1-2\Sigma_+)^2 + \Omega (R + R_+)](2-\Sigma_+)^{-1}\ .
\eeqa 

\noindent
{\bf Theorem 5.1} If a smooth non-vacuum reflection-symmetric LRS
solution of Bianchi type III of the Einstein-Vlasov equations for massive 
particles is represented as a solution of (\ref{eq:eq1}) with $M_2=0$,
then for $\tau \rightarrow \infty$ it converges to a point of the line 
$L_3$ with $z>0$.
For $\tau \rightarrow -\infty$ there exists 

\noindent
(i) a one-parameter set of solutions lying on the unstable manifold 
of $P_1$ and

\noindent 
(ii) a two-parameter set of solutions lying on the unstable manifold 
of $P_2$ and

\noindent
(iii) all remaining solutions converge to $P_3$.

\noindent
In all these solutions the scale factor $a$ is monotone increasing at late
times.

\noindent
{\bf Lemma 5.1} If a solution belongs to the interior of the type III state
space any $\alpha$-limit point satisfies $M_3=0$. Any $\omega$-limit point 
satisfies $\Sigma_+=\fr12$, $M_3=\fr34$ and $s=1$.

\noindent
{\bf Proof} The continuous function $\tilde M_3$ on the state space
must have a maximum and since its gradient never vanishes this maximum can 
only be attained at points with $M_3=1-\Sigma_+^2$. Computing the 
derivative of $\tilde M_3$ along the curve in the $(M_3,\Sigma_+)$ plane 
defined by this relation shows that the maximum value is $\fr13$ and that it 
is attained when $\Sigma_+=\fr12$ and $M_3=\fr34$. Now we apply the 
monotonicity principle. Let $S$ be the part of the 
Bianchi III state space obtained by removing the points with $M_3=0$ and 
those with $\Sigma_+-\fr12=M_3-\fr34=0$. This is an invariant set for the 
dynamical system. It will now be shown that $\tilde M_3$ is strictly 
increasing along solutions on this set. If $\Sigma_+\ne \fr12$ or if
$\Omega(R+R_+)\ne 0$ then this follows immediately from (\ref{eq:mon3}). 
If $\Sigma_+=\fr12$ and $\Omega(R+R_+)=0$ then
\beq\label{sigmadot}
\dot\Sigma_+=\fr34 (M_3-\fr34)
\eeq  
This completes the proof that $\tilde M_3$ is strictly increasing on $S$.
The monotonicity principle then shows that any point in the $\alpha$-limit
set must be in the complement of $S$ and such that such that $\tilde M_3$ 
does not take on its maximum value on $\bar S$ there. Hence $M_3=0$ there. 
It also shows that any point in the $\omega$-limit set must be in
the complement of $S$ and that $\tilde M_3$ does not take on its  
minimum value there. Hence in the latter case $\Sigma_+=\fr12$ and 
$M_3=\fr34$. It follows from this that $\Sigma_+\to \fr12$ as 
$\tau\to\infty$ and the equation for $s$ then implies that $s\to 1$.

\noindent
{\bf Lemma 5.2} A solution which belongs to the interior of the type 
III state space converges to a point of the line $L_3$ with $z>0$
as $\tau\to\infty$.

\noindent
{\bf Proof} Because of the result of Lemma 5.1 it only remains to prove
that $z$ tends to a positive limit  as $\tau\to\infty$. Note first that the 
evolution equation for $s$ implies an equation of the form 
$(d/d\tau)(1-s)=(1-s)F$ where $F=-6s\Sigma_+$. As $\tau$ tends to infinity 
$F\to -3$ and a simple 
comparison argument proves that $1-s(\tau)=O(e^{(-3+\epsilon)\tau})$ as 
$\tau\to\infty$. In particular, $1-s$ decays exponentially to zero at late
times. The evolution equations imply that $\Omega$ satisfies the equation:
\beq
\dot\Omega/\Omega=(\Sigma_+-\fr12)[(3-R)\Sigma_++\fr32 (1+R)]-2\Sigma_+
(R_++R)-(M_3-\fr34)(1+R)
\eeq
Note that $R_++R\ge 0$ so that the second term on the right hand side is
negative. However it is exponentially small at late times since it
contains a factor $(1-s)$ when expressed in terms of the matter quantities.
In particular $\dot\Omega/\Omega\to 0$ as $\tau\to\infty$ and 
$\Omega^{-1}=O(e^{\epsilon\tau})$ for any $\epsilon>0$. This means that 
$\Omega$
converges to zero slower than any exponential. In other words, 
$\Omega e^{\epsilon\tau}$ tends to infinity for any $\epsilon>0$.
Suppose that 
$\Sigma_+\ge \fr12$ for some solution at some time. Then $M_3\le \fr34$ and
the first and third terms in the expression for $\dot\Omega/\Omega$ are
positive at late times while the second term is negative. It will now be
shown that the third term decays slower than any exponential and thus must 
eventually dominate the second term. For
\beq
\Omega=(\fr14-\Sigma_+^2)+(\fr34-M_3)\le (\fr34-M_3)
\eeq
It follows that $\dot\Omega/\Omega>0$ at late times as long as 
$\Sigma_+>\fr12$. Since it follows from Lemma 5.3 that $\Omega\to 0$ as
$\tau\to\infty$ it follows that for any time $\tau_0$ for which 
$\Sigma_+(\tau_0)>\fr12$ there exists a time $\tau>\tau_0$ with 
$\Sigma_+=\fr12$. When $\Sigma_+=\fr12$ then
\beq\label{sigma12}
\dot\Sigma_+=(M_3-\fr34)[\fr34 (1-R)-\fr14 (R+R_+)]
\eeq 
Now it follows from the evolution equation for $z$ that $1-z$ cannot
approach zero faster than, for instance, $e^{-\tau}$ and the same is
then true of $1-R$. It can be concluded that the first term in the 
square bracket on the right hand side of (\ref{sigma12}) dominates
the second at late times. Hence $\fr12-\Sigma_+$ must be negative at late 
times, which in turns implies that $z$ is increasing and that it must
tend to a positive limit.

\noindent
{\bf Lemma 5.3} If a solution lies in the interior of the type III state
space then unless it lies on the unstable manifold of $P_1$ or $P_2$
(and both of these cases occur) the $\alpha$-limit set consists of the
point $P_3$.

\noindent
{\bf Proof} Note first that it can be concluded as in the proof of Lemma 3.4
that any $\alpha$-limit point satisfies $z=0$. Thus, applying Lemma 5.1, it 
can be identified with a point of the state space for massless type I 
solutions. Now it is possible to proceed further following the method of
proof of Lemma 3.4. Consider the boundary of the state space for massless
type I solutions. The point $P_3$, being a hyperbolic source in the type III
state space, can be excluded as an $\alpha$-limit point of a solution of 
type III. It is then possible to successively exclude points of the boundary
as in the proof of Lemma 3.4. The facts which need to be used are that all
$\alpha$-limit points satisfy $M_3=0$ and $z=1$ and that the points $P_4$,
$P_5$ and the endpoint of $L_1$ are a hyperbolic saddle, a non-hyperbolic
saddle topologically equivalent to a hyperbolic one and a transversely
hyperbolic saddle, respectively. At this stage it can be concluded that all
$\alpha$-limit points of solutions of type III are either $P_1$, $P_2$ or
points of the unstable manifold of $P_1$. For all other points of the 
interior of the massless type I state space lie on solutions which  
converge to the hyperbolic source $P_3$ in the past time direction, and so 
are excluded. It remains to examine what happens in a neighbourhood of 
the points $P_1$ and $P_2$, which are both hyperbolic saddles. The unstable 
manifold of $P_2$ in the type III state space is three-dimensional and so
there are solutions which converge to $P_2$ as $\tau\to -\infty$. Any other
type III solutions which had $P_2$ as an $\alpha$-limit point would have to
have $\alpha$-limit points on the stable manifold of $P_2$, which has 
already been excluded. Hence solutions of type III which do not converge 
in the past to $P_2$ cannot have $P_2$ or a point of its unstable manifold
as $\alpha$-limit points. Thus the only remaining possibility is that 
solutions lie on the unstable manifold of $P_1$ and converge to that point
in the past. Since the unstable manifold is two-dimensional, solutions of
this kind exist.   

\vskip 10pt\noindent
The results of Lemma 5.2 and Lemma 5.3 together imply all the results of
Theorem 5.1 except the last directly. The statement about the scale factor 
$a$ follows from the fact, derived in the course of the proof of Lemma 5.2,
that $\Sigma_+<\fr12$ at late times.

\section{Concluding remarks}

In this paper we studied the dynamics of solutions of the Einstein-Vlasov 
equations which are locally rotationally symmetric, reflection-symmetric
and of Bianchi types I, II and III. The initial singularities are of four
types. There are isotropic singularities which, in the dynamical systems
description used in this paper, are those which converge to the point 
$P_1$ as $\tau\to -\infty$. The general theory of isotropic singularities
developed by Anguige and Tod \cite{anguige99a, anguige99b} implies as a
very special case the occurrence of isotropic singularities in Bianchi
models with collisionless matter and information about how many there are.
They only developed the theory for massless particles and so in order to
apply to the situations considered here it would have to be generalized to
the massive case. There are barrel singularities which occur in types I and
III but not in type II.  In the dynamical systems picture these are the 
solutions which converge to $P_2$ as $\tau\to -\infty$. Fluid models with
corresponding symmetries never have barrel singularities and so this is a
peculiarity of collisionless matter, both in the case of massive 
particles studied here and that of massless particles studied in 
\cite{rentod99}. There is the generic case in types I and III, which
concerns solutions which develop from an open dense set of initial data
for each of these Bianchi types. These solutions have a cigar singularity
and converge to $P_3$ as $\tau\to -\infty$. Finally, there are the generic
solutions of type II, which have an oscillatory initial singularity.

As far as the late time behaviour is concerned, it is tempting to speculate
that behaving like a dust model at late times in an expanding phase may be
a general feature of solutions of the Einstein-Vlasov equations with 
massive particles. We know of no counterexample to this. For the solutions
of types I and II treated in this paper it has been proved to be true. For
type III the situation appears to be delicate and the occurrence of 
degenerate stationary points of the dynamical system may require an
application of centre manifold theory in order to determine details of
the asymptotics. A possible criterion for detecting cases where there may
be trouble is as follows. Consider a dust solution which is a candidate
for the asymptotic state of solutions of the Einstein-Vlasov equations. If
each eigenvalue of the second fundamental form of the homogeneous 
hypersurfaces, when divided by the mean curvature, is bounded below by a
positive constant in the dust solution then it is a strong candidate.
Otherwise difficulties are to be expected. This criterion gives a positive 
recommendation for types I and II and a warning for type III. Thus at least 
for the models investigated in this paper it is a good guide. Using the 
information on dust models in chapter 6 of \cite{waiell97} it also gives
a positive recommendation for types I, II and VI${}_0$ without the need
to restrict to the LRS case.

In this paper a dynamical system has been set up for all LRS Bianchi models
of class A
as well as for Kantowski-Sachs models and the type III models, which are of
class B. We expect that techniques similar to those used here can be applied 
to analyse Kantowski-Sachs models and LRS models of type VIII and IX. An 
important feature of all these LRS models is that the Vlasov equation can 
be solved exactly. This is also true of general Bianchi type I models. Some 
limited results on the dynamics of Bianchi type I solutions of the 
Einstein-Vlasov equations which are reflection-symmetric but not necessarily
LRS were proved in \cite{rendall96}. A heuristic analysis of 
reflection-symmetric type I models was given using Hamiltonian techniques 
in \cite{misner68}, where there are also interesting remarks on the general
Bianchi I case. It would be very desirable to have a mathematically rigorous 
implementation of the ideas of \cite{misner68}.

What can be done in cases where the Vlasov equation cannot be solved 
exactly? If, as already speculated above, the late time evolution resembles 
that of
a dust solution and if the dust solutions are asymptotically LRS then it
may be possible to give a good approximation to the solution of the 
Vlasov equation in that regime. There is one drawback of this idea as a
general tool for Bianchi class A models. Unfortunately there are no LRS 
spacetimes of Bianchi type VI${}_{0}$. In the case of fluids there 
exists a special class of Bianchi VI${}_{0}$ spacetimes which is often
characterized by the rather opaque statement that $n^\alpha_\alpha=0$.
These spacetimes do have a simple geometric characterization which will
now be explained. Every Bianchi class A spacetime has a discrete group
of isometries whose generators simultaneously reverse two of the 
invariant one-forms on the group. The special class of Bianchi VI${}_{0}$
solutions can be characterized by the existence of an additional isometry
which reverses just one of the one-forms. It is possible to consider
solutions of the Einstein-Vlasov equations with the corresponding type of
symmetry. We are, however, not aware that the Vlasov equation can be 
solved exactly in these special spacetimes. If it could then this might
fill the apparent gap in the strategy just suggested.

The oscillatory behaviour observed near the singularity in type II models
appears at first sight to indicate that collisionless matter does not fit 
into the analysis of general spacetime singularities by Belinskii,
Khalatnikov and Lifshitz \cite{bkl82}. On the other hand, the fact that 
in the analysis of Misner \cite{misner68} using a time-dependent potential
we see the phenomenon of walls moving too fast to be caught suggests that
the oscillations might go away in general models. This issue requires
further work. It could turn out that collisonless matter generically
becomes negligible near the singularity, as originally stated for fluids
in \cite{bkl82}.

To conclude, we mention some further interesting open problems. What happens
in the case of a model with two species of particles, one massive and one
massless? Of course this could be thought of as a simple cosmological model
incorporating both baryonic matter and the microwave background photons.
It is related to the two-fluid models which have been analysed in 
\cite{coley92}. Mixtures of fluids and kinetic theory could also be 
considered. We have seen that the Einstein equations with collisionless 
matter as source may behave very differently from the Einstein equations
with a fluid source at early times (and also at late times in the massless
case). Under what circumstances are there intermediate stages of the 
evolution with collisionless matter which can be well described by
a fluid? Since the point $P_1$ is a saddle there are obviously solutions
which approach this point and then go away again but is there more that
can be said about this issue? What can be said about inhomogeneous models?
In \cite{rein96} Rein analysed the behaviour at early times of solutions
of the Einstein-Vlasov equations with spherical, plane and hyperbolic 
symmetry and massive particles. He identified open subsets of initial
data for these symmetry types with a singularity resembling the generic 
LRS solutions of types I and III. There is an overlap between the results
of \cite{rein96} and those of the present paper. It could be 
illuminating to attempt a common generalization of these. In any case, it
is clear that one of the central challenges of the future in the study
of cosmological solutions of the Einstein-Vlasov equations, or indeed
the Einstein equations coupled to any type of matter fields, is to 
develop techniques which apply to inhomogeneous problems. A thorough 
understanding of the homogeneous case is likely to be an invaluable
guide in addressing it.

\appendix
\section{Appendix}

In this appendix some general procedures which are useful in determining 
limit sets of solutions of dynamical systems will be outlined. Let 
$\gamma$ be an orbit of a dynamical system and $p$ a stationary point.
We will discuss only $\omega$-limit sets, but corresponding statements
about $\alpha$-limit sets follow immediately by reversing the direction 
of time. We consider the following three statements which may or may not 
be true for given choices of $\gamma$ and $p$.

\begin{enumerate}
\item $p$ is an $\omega$-limit point of $\gamma$
\item $\gamma$ lies on the stable manifold of $p$
\item there are $\omega$-limit points of $\gamma$ different from $p$
which are arbitrarily close to $p$ and lie on the unstable manifold of $p$
\item there are $\omega$-limit points of $\gamma$ different from $p$
which are arbitrarily close to $p$ and lie on the stable manifold of $p$
\end{enumerate} 

In the body of the paper we frequently use certain relations among the 
statements above which hold under various assumptions on the nature of
the stationary point $p$. Whatever the stationary point, it is always 
true that the statement 1. is implied by any of the statements 2., 3. or 4. 
This is a consequence of the elementary fact that the $\omega$-limit set 
is closed. Now suppose that $p$ is a hyperbolic stationary point. In this 
case, if 1. is true and 2. is false then both 3. and 4. are true. This
follows from Lemma A1 of \cite{rentod99}. Combining these statements we
see that for a hyperbolic stationary point there are two mutually exclusive 
cases under which 1. can hold. Either $\gamma$ lies on the unstable manifold
of $p$ or the $\omega$-limit set contains points of both the stable and 
unstable manifolds of $p$ arbitrarily close to $p$. In particular, if $p$
is a hyperbolic source then it cannot be in the $\omega$-limit set of 
$\gamma$ and if $p$ is a hyperbolic sink and $p$ is in the $\omega$-limit
set of $\gamma$ it is the whole $\omega$-limit set. If we already have some
a priori information about where $\omega$-limit points can lie (due, for
instance, to the existence of a monotone function) then this gives more 
information about where the points on the stable and unstable manifolds
whose existence is guaranteed by the general statements above can lie.

Next we consider the case of transversally hyperbolic stationary points.
Suppose that $p$ belongs to a manifold of stationary points of dimension
$d$. (Only the case $d=1$ occurs in this paper.) These points have a 
zero eigenvalue of multiplicity $d$. If all other eigenvalues have 
non-vanishing real parts then the stationary point is called 
transversally hyperbolic. (Depending on the signs of the eigenvalues the
manifold of stationary points
is called a transversally hyperbolic source, sink or saddle.)
By the reduction theorem (Theorem A1 of 
\cite{rentod99}) each of these points lies on an invariant manifold
and the restriction of the flow to each invariant manifold is 
topologically equivalent to that near a hyperbolic stationary point.
The arguments for a hyperbolic stationary point adapt easily to give
analogous statements for transversally hyperbolic stationary points.
In applying these results we can essentially ignore the directions
along the manifold of stationary points.

Finally we consider certain other non-hyperbolic stationary points. A 
result of the type we need was proved in Lemma A2 of \cite{rentod99} but 
we would like to formulate the statement in a more transparent way here.
Consider an isolated stationary point $p$ with a trivial stable manifold 
and a one-dimensional centre manifold. Using the reduction theorem we see 
that the unstable manifold divides a neighbourhood of $p$ into two parts
on each of which the restriction of the dynamical system is 
topologically equivalent to the restriction of a dynamical system with
a hyperbolic stationary point. Whether the latter system has a saddle
or a source depends on a certain sign condition. This condition may be
different for the two halves. In the dynamical systems considered in
this paper the only example of this is provided by the point $P_5$. Only
one of the halves belongs to the physical part of the state space and in 
that half the sign is such that a saddle is obtained. The result of these
considerations is that for the arguments in this paper $P_5$ may be 
treated just as if it had been a hyperbolic saddle, with the centre 
manifold taking over the role of the trivial stable manifold.



\begin{thebibliography}{00}

\bibitem{anguige99b} Anguige K 1999 Isotropic cosmological singularities 3: 
The Cauchy problem for the inhomogeneous conformal Einstein-Vlasov 
equations. Preprint gr-qc/9903018.

\bibitem{anguige99a} Anguige K and Tod K P 1999 Isotropic cosmological 
singularities 2: The Einstein-Vlasov system. {\it Ann. Phys. (NY)} {\bf 276} 
294-320 

\bibitem{ash93}
Ashtekar A, Tate R S and Uggla  C 1993 Minisuperspaces: observables and
quantization. {\it Int. J. Mod. Phys. D} {\bf 2} 15-50

\bibitem{bkl82} Belinskii V A, Khalatnikov I M and Lifshitz E M 1982 A 
general solution of the Einstein equations with a time singularity.
{\it Adv. Phys.} {\bf 31} 639-67

\bibitem{coley92} Coley A A and Wainwright J 1992 Qualitative analysis
of two-fluid Bianchi cosmologies. \CQG {\bf 9} 651-665

\bibitem{jan84}
Jantzen R T 1987
{\it Proc.\ Int.\ Sch.\ Phys.\ `E Fermi' Course LXXXVI on `Gamov
  Cosmology'\/}
ed R Ruffini and F Melchiorri (Amsterdam: North Holland); 
1984
{\it Cosmology of the Early Universe\/} 
ed L Z  Fang and R  Ruffini 
(Singapore:  World Scientific) 

\bibitem{misner68} Misner C W 1968 The isotropy of the universe. 
{\it Astrophysical Journal} {\bf 151} 431-457

\bibitem{rein96} Rein G 1996 Cosmological solutions of the Vlasov-Einstein
system with spherical, plane or hyperbolic symmetry. {\it Math. Proc. Camb.
Phil. Soc.} {\bf 119} 739-762

\bibitem{rendall96} Rendall A D 1996 The initial singularity in solutions of 
the Einstein-Vlasov system of Bianchi type I. {\it J. Math. Phys.} {\bf 37}  
438-451

\bibitem{rentod99}
Rendall A D and Tod K P 1999
Dynamics of spatially homogeneous solutions of the 
Einstein-Vlasov equations which are locally rotationally symmetric.
\CQG {\bf 16}  1705-1726

\bibitem{uggzur90}
Uggla C and von Zur-M\"uhlen H 1990
Compactified and reduced dynamics for locally rotationally 
symmetric Bianchi type IX perfect fluid models.
\CQG {\bf 7} 1365-1385

\bibitem{uggjanros95}
Uggla  C, Jantzen R T and Rosquist K 1995
Exact hypersurface-homogeneous solutions in cosmology and astrophysics.
{\it Phys. Rev. D} {\bf 51}  5522-5557. 

\bibitem{waiell97}
Wainwright J and Ellis G F R 1997
{\it Dynamical Systems in Cosmology\/}
(Cambridge: Cambridge University Press)

\end{thebibliography}
\end{document}